\documentclass[aps,prl,twocolumn,superscriptaddress]{revtex4}
\usepackage{graphicx}% Include figure files
\usepackage{dcolumn}% Align table columns on decimal point
\usepackage{bm}% boldmath \citestyle{prb}

\begin{document}

\title{Microscopic structure and dynamics of high and low density trans-1,2-dichloroethylene liquids}

\author{M. \surname{Rovira-Esteva}}
\affiliation{Grup de Caracteritzaci{\'o} de Materials,
Departament de F\'{\i}sica i Enginyeria Nuclear, ETSEIB,
Universitat Polit{\`e}cnica de Catalunya, Diagonal 647, 08028
Barcelona, Catalonia, Spain}

\author{A. Murugan}
\affiliation{Grup de Caracteritzaci{\'o} de Materials,
Departament de F\'{\i}sica i Enginyeria Nuclear, ETSEIB,
Universitat Polit{\`e}cnica de Catalunya, Diagonal 647, 08028
Barcelona, Catalonia, Spain}

\author{L. C. Pardo}
\affiliation{Grup de Caracteritzaci{\'o} de Materials,
Departament de F\'{\i}sica i Enginyeria Nuclear, ETSEIB,
Universitat Polit{\`e}cnica de Catalunya, Diagonal 647, 08028
Barcelona, Catalonia, Spain}

\author{S. Busch}
\affiliation{Physik Department E13 and
Forschungsneutronenquelle Heinz Maier-Leibnitz (FRM II), Technische
Universit\"at M\"unchen, Lichtenbergstr.\ 1, 85748 Garching,
Germany}

\author{M. D. \surname{Ruiz-Mart{\'i}n}}
\affiliation{Grup de Caracteritzaci{\'o} de Materials,
Departament de F\'{\i}sica i Enginyeria Nuclear, ETSEIB,
Universitat Polit{\`e}cnica de Catalunya, Diagonal 647, 08028
Barcelona, Catalonia, Spain}

\author{M.-S. Appavou}
\affiliation{Forschungszentrum J\"ulich GmbH, Institut f\"ur
Festk\"orperforschung (IFF), J\"ulich Centre for Neutron Science
(JCNS), FRM II outstation, Lichtenbergstr.\ 1, 85748 Garching,
Germany}

\author{J. Ll. Tamarit}
\affiliation{Grup de Caracteritzaci{\'o} de Materials,
Departament de F\'{\i}sica i Enginyeria Nuclear, ETSEIB,
Universitat Polit{\`e}cnica de Catalunya, Diagonal 647, 08028
Barcelona, Catalonia, Spain}

\author{C. Smuda}
\affiliation{ETH Z\"urich, Center for Radiopharmaceutical Science,
 Wolfgang-Pauli-Str. 10, CH-8093 Z\"urich, Switzerland}

\author{T. Unruh}
\affiliation{Physik Department E13 and
Forschungsneutronenquelle Heinz Maier-Leibnitz (FRM II), Technische
Universit\"at M\"unchen, Lichtenbergstr.\ 1, 85748 Garching,
Germany}

\author{F. J. Bermejo}
\affiliation{Facultad de Ciencia y Tecnolog\'{\i}a, Universidad del Pa\'{\i}s Vasco / EHU, P.
Box 644, E-48080 Bilbao, Spain}

\author{G. J. Cuello}
\affiliation{Facultad de Ciencia y Tecnolog\'{\i}a, Universidad del Pa\'{\i}s Vasco / EHU, P.
Box 644, E-48080 Bilbao, Spain}
\affiliation{Institut Laue Langevin, 6 Rue Jules Horowitz, BP.
156, F-38042 Grenoble Cedex 9, France}

\author{S. J. Rzoska}
\affiliation{Institute of Physics, Silesian University, Uniwersytecka 4, 40-007 Katowice, Poland}

\begin{abstract}
We present a study of the dynamics and structural changes for trans-1,2-dichloroethylene between high and low density liquids
using neutron scattering techniques (diffraction, small angle neutron scattering and
time of flight spectroscopy) and molecular dynamics simulations. We show that changes in the short range
ordering of molecules goes along with a change of the molecular dynamics: both structure and dynamics of the high density liquid are more cooperative than those of the low density liquid. The microscopic mechanism underlying the cooperative motions in the high density liquid has been found to be related to the backscattering of molecules due to a strong correlation of molecular ordering.
\end{abstract}


\maketitle

Classical thermodynamics establishes the existence of one unique
liquid state for any material, i.e., there is only a liquid phases characterized by its density given the thermodybamic coordinates pressure and temperature. Nevertheless, recent experimental
results and molecular dynamics (MD) simulations suggest that,
even for one-component systems, several liquid phases can appear
with an associated liquid-liquid phase transition (LLPT). A noticeable
number of cases has been found for atomic liquids,
the best-known example concerning liquid phosphorus \cite{fosforohp,fosforohp2},
where the LLPT appears as a transition between thermodynamically stable phases with strong structural changes \cite{fosforo}.
As far as molecular liquids are concerned, the number of experimental evidences for LLPT is still rather
scarce and comprises only a limited number of compounds such as
triphenyl phosphite \cite{TPP} and n-butanol \cite{butanol}. According to the so called two order parameter theories that propose an explanation for the LLPT, liquids must be described not only by their density but also by an additional order parameter accounting for changes in the molecular arrangement \cite{TOP,TOP2,TOP3}.
The LLPT can end in a liquid-liquid critical point between a high-density liquid (HDL) and a low-density liquid (LDL). However, changes in the dynamics with an associated change in structural features can also be explained by a singularity-free scenario \cite{franzesedyn}. In the latter case, changes in both dynamics and structure from a HDL to a LDL also take place at the point where the isobaric heat capacity $C_P$ has a maximum, but no critical point or LLPT are observed at non-zero temperature.

An early work on trans-1,2-dichloroethylene ($T_{\text{melt}}=223~\textrm{K}$) suggested the existence of a LLPT at $T_t=247~K > T_{\text{melt}}$ based on a small jump in density (less than 0.06\%) as well as in the compressibility, and a clear discontinuity on the spin-lattice relaxation time $T_1$ \cite{kawanishi,kawanishi2}. The observed changes in $T_1$ were tentatively related to a lack of freedom of the molecular rotation in the HDL, not present in the LDL. The change on the dynamics of this substance between both liquids, was thereafter also supported by discontinuities in the viscosity measurements and the slope of the rotational relaxation time \cite{raman}, and by the absorbance, frequency and linewidth of several infrared vibrational spectroscopy bands \cite{infraroig}. Concerning structural related magnitudes, subsequent measurements of the density as a function of the temperature
did not find a jump at the expected LLPT but only a change in its slope \cite{raman}. More recently, some of the authors of the present work have also undertaken calorimetric and nonlinear dielectric experiments \cite{rzoska}. In that work a strong pre-transitional anomaly of nonlinear dielectric effect was obtained, similar to the one observed in the isotropic phase of nematic liquid crystals, together with a maximum of $C_P$. Therefore, experimental results unambiguously show a clear change in the dynamics, together with a slight change in the structure, between HDL and LDL, that takes place when $C_p$ is in a maximum. However, those facts are not enough to unambiguously determine if they are related to a singularity-free or a liquid-liquid critical point scenario \cite{franzesedyn}, i.e., the liquid undergoes a first order phase transition. The present study is aimed to investigate the microscopic structural and dynamical differences between the HDL and LDL, from the experimental point of view and from MD simulations.

\begin{figure} \includegraphics[width=8.2cm]{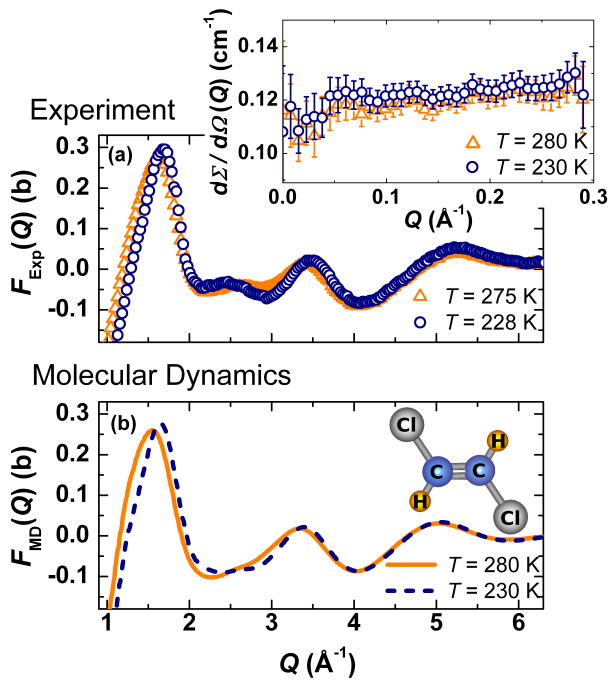}
\caption{\label{fig1}(color online). Total interference function of
trans-1,2-dichloroethylene for the LDL and the HDL, obtained by neutron scattering
experiments (a) and by molecular dynamics simulations (b), where an ab initio calculation of the molecular structure is also shown. Inset shows the macroscopic cross section for low \textit{Q} values obtained by
SANS experiments also for temperatures above and below $T_t$.}
\end{figure}
Because the differences between the HDL and LDL in trans-1,2-dichloroethylene were related to a change
from a nematic-like to an isotropic molecular ordering as temperature is raised \cite{kawanishi,kawanishi2},
we have performed a series of small angle neutron scattering (SANS) measurements from 220 K to room temperature
to ascertain whether there is formation of intermediate range nematic-like structures or clustering in the HDL. Experiments were
performed on the KWS-2 diffractometer of the J\"ulich Centre for Neutron Science at the Forschungsneutronenquelle Heinz Maier-Leibnitz (FRM II, Munich, Germany) \cite{KWS2} using a wavelength of 4.5~\AA\ and a sample-detector distance of 2.0~m that allowed to perform measurements in the \textit{Q} range between $0.003$ and $0.3~\text{\AA}^{-1}$. The program \textsc{qtikws} \cite{QtiKWS} was used to perform data correction and normalization.
Results are shown in the inset of Fig.\ \ref{fig1} for the HDL and LDL. No pronounced signal has been obtained within the experimental error in the measured \textit{Q}-range, which disfavors the existence of a long range nematic-like ordering for the HDL \cite{liquidcrystals}.
The data exclude that the differences between both liquids are related to the emergence of molecular clustering on length scales of about 20--2000 \AA.

Going down in the spatial range, the microscopic short range order (SRO) concerning a length scale of the order of a few molecular lengths ($l_{\text{mol}}\approx 4~\text{\AA}$) has been analyzed by means of neutron scattering experiments on the D4c diffractometer at the Institute Laue-Langevin (ILL, Grenoble, France) \cite{d4c} using a wavelength of 0.5~\AA\ and a deuterated sample. Data were corrected and normalized using the program \textsc{correct} \cite{correct} and inelastic corrections were also carried out (for details on data reduction see Ref.\ \cite{CCl4}).
The obtained total interference function $F(Q)$ \cite{barnes} is shown in Fig.\ \ref{fig1}  for two representative temperatures. A change in the shape of the profile emerges between the first and second peak which reflects a change in the SRO.
This change was also observed in a series of temperature dependent experiments on the D20 diffractometer (ILL) using a wavelength of 1.3~\AA, giving better access to the low-$Q$ region. It should be pointed out that similar changes have been found in the case of experiments performed on HDL water at high pressures \cite{thierrywater} and MD simulations on HDL silicon \cite{sastry,sastry2}.

The microscopic mechanisms giving rise to the changes in the
interference function $F(Q)$ have been investigated through a
series of MD simulations (Fig.\ \ref{fig1}). Those were performed using the program \textsc{amber8} \cite{GAAF,GAAF2} with a simulation box containing 3629 molecules and a time step of 1~fs. The inter- and intra-molecular interactions for the trans-1,2-dichloroethylene molecule
were described using the GAFF force field \cite{GAAF,GAAF2}. The simulations were carried out for the temperature
range 200--350~K in the NPT ensemble, therefore allowing the box size to change, and the total time scale of each simulation run was 40--50 ns \cite{arul,arul2}.
As can be seen in Fig.\ \ref{fig1}, the agreement between simulations and
experiment is excellent. Note that not only the shapes of the simulated $F(Q)$ closely resemble the experimental ones, but also positions and intensities are equal within the experimental error.

\begin{figure} \includegraphics[width=8.5cm]{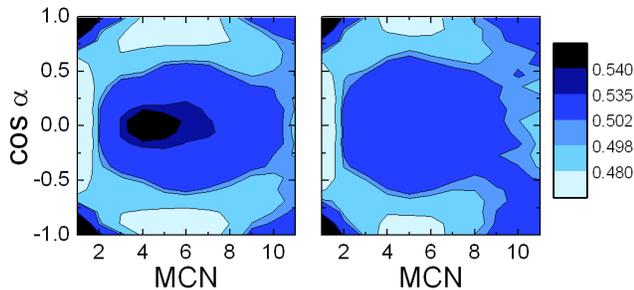}
\caption{\label{fig2}(color online). Short range order for the HDL (left) and the LDL (right), at 200 and 300 K respectively. The \textit{y}-axis is the cosine of the angle $\alpha$ formed between the vectors defined by the C--C double bonds of two neighbor molecules and the \textit{x}-axis is the number of molecules surrounding a central one. The \textit{z}-axis represents the probability of finding two molecules at a distance determined by the MCN with a relative orientation defined by the $\alpha$ angle.}
\end{figure}

To emphasize quantitatively which are the SRO changes between the HDL and LDL, we show in  Fig.\ \ref{fig2} the probability of finding two molecules with a determined relative orientation for increasing distances. To present the distance dependent SRO, the molecular coordination number (MCN) has been chosen rather than distance in order to avoid trivial effects due to density changes. Figure \ref{fig2} shows that the relative
orientation of nearest neighbors is virtually the same and strongly defined for both liquids, in such a way that their C--C vectors are parallel, i.e., with $\cos \alpha=\pm1$. However, upon increasing
distance between the molecules (third to tenth neighbor), a clear difference between the LDL and HDL can be seen in the SRO. For instance, for MCN between eight and ten, molecules are arranged in a  orthogonal way for HDL and are randomly oriented for the LDL. Further analysis (not shown) tell us that besides the relative orientation of the C--C axes of two molecules, the reported changes are also reflected in the relative position of two molecules and in the relative orientation of the planes
defined by the Cl--C--H bonds of two molecules. As proposed in the frame of two order parameter theories
\cite{TOP,TOP2,TOP3,franzesedyn}, this liquid can therefore not be simply
characterized by its density, but also a parameter reflecting the
SRO should account for the changes occurring between the two liquids.

\begin{figure} \includegraphics[width=8.3cm]{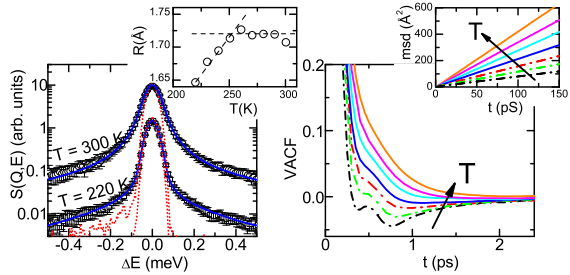}
\caption{\label{fig3}(color online). (a) Fits of QENS spectra for $Q = 0.3 ~\text{\AA}^{-1}$ at 300 K (shifted upwards) to a model considering diffusion and rotation motions, and to a model with an additional
confined motion for 220 K (solid lines). Dotted lines show the experimental resolution. Inset shows radius of rotation in function of temperature for the model with only diffusion and rotation (dashed lines are a guide to the eye). (b) VACF obtained by MD simulations as a function of time, for temperatures from 200 to 320~K in steps of 20 K (solid lines for the LDL and dashed lines for the HDL). The inset shows the obtained mean square displacement from simulations (solid lines for the LDL and dashed lines for the HDL).}
\end{figure}

To ascertain the influence of the aforementioned
structural changes in the dynamics of the system, we have
performed a series of quasielastic neutron scattering (QENS) experiments for temperatures ranging from 220 to 300 K, conducted with the TOFTOF spectrometer at FRM II on a hydrogenated sample. Spectra were measured using an energy resolution of $60~\mu\text{eV}$ and a wavelength of 6~\AA, and data reduction was performed using the program \textsc{frida} \cite{software}. Two representative spectra at $Q=0.3~\text{\AA}^{-1}$ at the LDL and HDL are
shown in Fig.\ \ref{fig3}(a).

To have a first insight of possible changes in the dynamics, a stretched exponential $\exp(-t/\tau)^\beta$ was fitted to the intermediate scattering function, obtaining a decrease of the exponent $\beta$ in the HDL, which is related to a broadening of the relaxation time distribution, i.e., an increase of the cooperativity of the molecular motion. A deeper analysis was performed through a careful fit of data to several models using a Bayesian
approach with the program \textsc{fabada} \cite{bayes,bayes2,software2}. In order to keep the number of physical
parameters describing the data to a minimum, we performed the
fits to the whole scattering law $S(Q,\omega)$. Model
selection was performed calculating the maximum of the
likelihood $\mathcal{L}^{\text{max}}$ for each model.

The first model used to describe the data is composed by a diffusion
motion plus an isotropic rotation of the molecule \cite{bee}.
In this way, the only physical parameters to describe the whole
experimental scattering function were the translational and rotational diffusion coefficients as well as the
radius of rotation $R$. This simple model is able to describe LDL data giving rise to a good quality fit and a radius of rotation almost independent of temperature ($R\approx 1.72~\text{\AA}$, see inset in Fig.\ \ref{fig3}(a)), in agreement with the aforementioned NMR measurements \cite{kawanishi,kawanishi2}. For the HDL an inability of this model to describe experimental data (on quantitative grounds, a decrease on the $\mathcal{L}^{\text{max}}$ of the fit), makes the isotropic model for rotation not valid to describe molecular rotation. This is also reflected on the sudden drop of \textit{R} below $T_t$ (inset in Fig.\ \ref{fig3}(a)). Even when assuming a free diffusion model for rotations \cite{egelstaff} and assuming an anisotropic rotation of molecules \cite{arotation,arotation2}, models could not account for spectra obtained in the HDL. Only adding a confined motion to the previous model, data could be successfully described. For the latter model, $S(Q,\omega)$ can be expressed as:
\begin{eqnarray*} \label{model}
S(Q,\omega)&=& \left[ A(Q) \delta( \omega )+(1-A(Q))\cdot
L(\omega)\right]  \\
 &\otimes&
 S_{\text{rot}}(Q,\omega) \otimes
S_{\text{diff}}(Q,\omega) \otimes R(Q,\omega)
\end{eqnarray*}
where $A(Q)$ is the elastic incoherent structure factor of
the confined motion, $L(\omega)$ is a Lorentzian function accounting for a confined motion, $S_{\text{rot}}(Q,\omega)$ is the component accounting for the molecular rotation, $S_{\text{diff}}(Q,\omega)$ the component for molecular diffusion \cite{bee}, and $R(Q,w)$ is the instrumental resolution.
Agreement with HDL spectra is shown in Fig.\ \ref{fig3}(a).
To estimate the length scale at which
the confined motion is taking place, $A(Q)$ values have been
fitted with a model of diffussion inside a sphere, yielding
$R=1.91 \pm 0.07 ~\text{\AA}$ and, more realistically, a threedimensional Brownian oscillator \cite{doster},
yielding a mean squared displacement $\sqrt{\delta^2}=2.12 \pm 0.07 ~\text{\AA}$. Moreover, the obtained $A(Q)$ are almost independent of temperature for the HDL, which means that the
length scale of the confined motion is roughly temperature independent for this liquid.

As previously performed for the microscopic structure investigation, the dynamics obtained from MD simulations have been analyzed as well. Agreement between translational diffusion activation energy determined using the
neutron experiments ($E_A=76\pm 1~\text{meV}$) and MD simulations
($E_A=75\pm 2~\text{meV}$) confirms that the simulation is indeed describing the dynamics of our system, and that we are actually simulating the liquid phase in all temperature range (see also inset of figure 3b) . The
normalized velocity autocorrelation function (VACF) is shown in Fig.\ \ref{fig3}(b). The fastest decay for the HDL is higher that that of LDL indicating that interaction with neighboring molecules takes before in that phase. As can be seen, there is a clear change in the dynamics of the two liquids, the VACF reaching negative values for HDL. Although interactions between
neighbor molecules at the LDL seem not to impede molecular diffusion, for the HDL a well defined backscattering effect
emerges, displaying a VACF with two minima characteristic to that of hydrogen bonded
systems \cite{sese}. Such a change in the dynamics is not
expected to be due to temperature effects \cite{dynamicliquids}.
Additionally, a change of density of the system would not be able to explain this
change in the dynamics since a change as large as about 15\% in density is needed to produce an effect on the liquid
dynamics in other systems \cite{dynamicliquids}. Therefore, only a change in the SRO is able to account
for the change in the dynamics seen by experiments and simulations.

The need for an additional confined motion at low temperatures to describe the QENS data goes along with a backscattering effect observed in the simulations. We can therefore assert that a cooperative molecular motion is present for the HDL but not for the LDL, which agrees with other experimental results \cite{infraroig}. The complex dynamics of the HDL agrees with a stronger molecular ordering present in this liquid, where on average there are many orthogonally oriented molecular pairs. This strong correlation is partially lost in the LDL, where molecular movements are due to non-coupled movements of diffusion and rotation.

We have shown that the microscopic ordering
of molecules and molecular dynamics are different between the HDL and the LDL. The changes in the dynamics are not due to
temperature or density effects, but due to changes in the SRO: for the HDL there are molecules perpendicularly oriented that are randomly oriented in the LDL. This accounts for the change of molecular dynamics from simple non-cooperative motions in the LDL to cooperative motions for the HDL. However, if these effects are to be explained in the frame of a singularity-free scenario or a liquid-liquid critical point scenario associated to a first order phase transition remains an open question.

\begin{acknowledgments}
The authors would like to thank C.~A. Angell, G. Franzese, G. Ses{\'e} and T. Str{\"a}ssle for helpful discussions and X. Ariza for deuterating the sample. This work has been supported by the Spanish Ministry of Science and Technology (FIS2008-00837, BES-2007-17418), by the Government of Catalonia (2009SGR-1251) and by the European Commission (NMI3/FP7).
\end{acknowledgments}
\end{document}